\documentclass[aps,prl,twocolumn,superscriptaddress,floats]{revtex4}

\usepackage[]{graphicx}
\usepackage[usenames,dvipsnames]{color}
\usepackage{soul}
\usepackage{bm}
\usepackage{amsmath}
\usepackage{txfonts}
\usepackage{amssymb}
\usepackage{graphicx}
\usepackage{multirow}
\usepackage{float}
\usepackage{indentfirst}

\begin{document}

\title{Approaching itinerant magnetic quantum criticality through a Hund's coupling induced electronic crossover in the YFe$_2$Ge$_2$ superconductor}
\author{D. Zhao}\affiliation{Hefei National Laboratory for Physical Science at Microscale and Department of Physics, University of Science and Technology of China, Hefei, Anhui 230026, People's Republic of China}
\author{H. L. Wo}\affiliation{State Key Laboratory of Surface Physics and Department of Physics, Fudan University, Shanghai 200433, China}
\author{J. Li}\affiliation{Hefei National Laboratory for Physical Science at Microscale and Department of Physics, University of Science and Technology of China, Hefei, Anhui 230026, People's Republic of China}
\author{D. W. Song}\affiliation{Hefei National Laboratory for Physical Science at Microscale and Department of Physics, University of Science and Technology of China, Hefei, Anhui 230026, People's Republic of China}
\author{L. X. Zheng}\affiliation{Hefei National Laboratory for Physical Science at Microscale and Department of Physics, University of Science and Technology of China, Hefei, Anhui 230026, People's Republic of China}
\author{S. J. Li}\affiliation{Hefei National Laboratory for Physical Science at Microscale and Department of Physics, University of Science and Technology of China, Hefei, Anhui 230026, People's Republic of China}
\author{L. P. Nie}\affiliation{Hefei National Laboratory for Physical Science at Microscale and Department of Physics, University of Science and Technology of China, Hefei, Anhui 230026, People's Republic of China}
\author{X. G. Luo}\affiliation{Hefei National Laboratory for Physical Science at Microscale and Department of Physics, University of Science and Technology of China, Hefei, Anhui 230026, People's Republic of China}\affiliation{Key Laboratory of Strongly-coupled Quantum Matter Physics, Chinese Academy of Sciences, University of Science and Technology of China, Hefei 230026, China}\affiliation{CAS Center for Excellence in Superconducting Electronics (CENSE), Shanghai 200050, China}\affiliation{CAS Center for Excellence in Quantum Information and Quantum Physics, Hefei, Anhui 230026, China}\affiliation{Collaborative Innovation Center of Advanced Microstructures, Nanjing 210093, China}
\author{J. Zhao}\affiliation{State Key Laboratory of Surface Physics and Department of Physics, Fudan University, Shanghai 200433, China}\affiliation{Collaborative Innovation Center of Advanced Microstructures, Nanjing 210093, China}
\author{T. Wu}\email{wutao@ustc.edu.cn}\affiliation{Hefei National Laboratory for Physical Science at Microscale and Department of Physics, University of Science and Technology of China, Hefei, Anhui 230026, People's Republic of China}\affiliation{Key Laboratory of Strongly-coupled Quantum Matter Physics, Chinese Academy of Sciences, University of Science and Technology of China, Hefei 230026, China}\affiliation{CAS Center for Excellence in Superconducting Electronics (CENSE), Shanghai 200050, China}\affiliation{CAS Center for Excellence in Quantum Information and Quantum Physics, Hefei, Anhui 230026, China}\affiliation{Collaborative Innovation Center of Advanced Microstructures, Nanjing 210093, China}
\author{X. H. Chen}\affiliation{Hefei National Laboratory for Physical Science at Microscale and Department of Physics, University of Science and Technology of China, Hefei, Anhui 230026, People's Republic of China}\affiliation{Key Laboratory of Strongly-coupled Quantum Matter Physics, Chinese Academy of Sciences, University of Science and Technology of China, Hefei 230026, China}\affiliation{CAS Center for Excellence in Superconducting Electronics (CENSE), Shanghai 200050, China}\affiliation{CAS Center for Excellence in Quantum Information and Quantum Physics, Hefei, Anhui 230026, China}\affiliation{Collaborative Innovation Center of Advanced Microstructures, Nanjing 210093, China}
\date{\today}

\begin{abstract}

Here, by conducting a systematic $^{89}$Y NMR study, we explore the nature of the magnetic ground state in a newly discovered iron-based superconductor YFe$_2$Ge$_2$. An incoherent-to-coherent crossover due to the Hund's coupling induced electronic correlation is revealed below the crossover temperature $T^*\sim 75\pm15\,\mathrm{K}$. During the electronic crossover, both the Knight shift ($K$) and the bulk magnetic susceptibility ($\chi$) exhibit a similar nonmonotonic temperature dependence, and a so-called Knight shift anomaly is also revealed by a careful $K$-$\chi$ analysis. Such an electronic crossover has been also observed in heavily hole-doped pnictide superconductors \emph{A}Fe$_2$As$_2$ (\emph{A} = K, Rb, and Cs), which is ascribed to the Hund's coupling induced electronic correlation. Below $T^*$, the spin-lattice relaxation rate divided by temperature $(1/T_1T)$ shows a similar suppression as the Knight shift, suggesting the absence of critical spin fluctuations. This seems to be in conflict with a predicted magnetic quantum critical point (QCP) near this system. However, considering a $\mathbf{q}$-dependent ``filter" effect on the transferred hyperfine field, a predominant spin fluctuation with A-type correlation would be perfectly filtered out at $^{89}$Y sites, which is consistent with the recent inelastic neutron scattering results. Therefore, our results confirm that, through a Hund's coupling induced electronic crossover, the magnetic ground state of YFe$_2$Ge$_2$ becomes close to an itinerant magnetic QCP with A-type spin fluctuations. In addition, the possible superconducting pairing due to spin fluctuations is also discussed.
\end{abstract}

\maketitle
\section{I. INTRODUCTION}
Superconductivity nearby spin order is always believed to have an unconventional pairing mechanism beyond electron-phonon interactions, such as cuprate superconductors, heavy-fermion superconductors, and iron-based superconductors (FeSCs)\,\cite{wen,Dai,heavy fermion,puzzle}. Spin fluctuation is a popular candidate for gluing electrons into Cooper pairs\,\cite{Scalapino}. Usually, antiferromagnetic (AFM) spin fluctuations favor spin-singlet pairing and ferromagnetic (FM) spin fluctuations favor spin-triplet pairing. In FeSCs, the stripe-type AFM spin fluctuations have been widely observed\,\cite{Dai}, which promotes early theory with spin-singlet pairing.
Recently, an indirect evidence for the coexistence of AFM and FM spin fluctuations was revealed by nuclear magnetic resonance (NMR) experiments in FeSCs with 122 structure\,\cite{NMR_FM1,NMR_FM2,NMR_FM3}.
It is found that the FM spin fluctuations are strongest in the maximally electron- and hole-doped BaCo$_2$As$_2$ and KFe$_2$As$_2$\,\cite{NMR_FM2}. This strongly suggests that the competition between AFM and FM spin fluctuations is a crucial ingredient to understand the variability of superconducting temperature $(T_c)$\,\cite{NMR_FM2}, especially for the domelike behavior. However, the direct evidence for FM spin fluctuations from polarized inelastic neutron scattering (INS) experiments is still absent\,\cite{neutron}. This prevents further understanding of the correlation between AFM and FM spin fluctuations in FeSCs.

The recent progress on bulk superconductivity in iron germanide compound YFe$_2$Ge$_2$ with $T_c$ below 1.8 K shed light on the above issue\,\cite{Y.zou,YFe2Ge_super}.
YFe$_2$Ge$_2$ has the same crystal structure as the 122-structure family of FeSCs and is isoelectronic to the maximally hole-doped KFe$_2$As$_2$, as shown in the inset of Fig. 1.
Due to the existence of Ge-Ge bonds, the Fermi surface geometry of YFe$_2$Ge$_2$ resembles that of KFe$_2$As$_2$ under high pressure\,\cite{Subedi,Singh}, which has a similar collapsed tetragonal phase (CTP) as the existence of As-As bonds. All these facts suggest that YFe$_2$Ge$_2$ is a good reference compound of KFe$_2$As$_2$ to investigate the correlation between AFM and FM spin fluctuations.
Theoretically, the standard density functional theory (DFT) calculation predicted that the magnetic ground state of YFe$_2$Ge$_2$ is an A-type order with a dominated in-plane ferromagnetic correlation\,\cite{Singh}. However, experimentally, there is no evidence for such magnetic order in YFe$_2$Ge$_2$\,\cite{Y.zou,YFe2Ge_super,YFe2Ge2 susci}. Only a large fluctuating magnetic moment on the Fe atom was revealed by the x-ray spectroscopy experiment\,\cite{YFe2Ge2 fluctuation moment}. The recent NMR experiment on YFe$_2$Ge$_{2-x}$Si$_x$ polycrystalline samples also supported the existence of FM spin fluctuations\,\cite{YFe2Ge2_NMR}. These findings support that YFe$_2$Ge$_2$ is close to a magnetic quantum critical point (QCP) with a predominant in-plane ferromagnetic correlation\,\cite{Singh}. Very recently, unambiguous evidence for the coexistence of stripe-type and A-type (in-plane FM correlation) spin fluctuations from INS experiments has been successfully found in YFe$_2$Ge$_2$ single crystals\,\cite{neutron}. The A-type spin fluctuations were enhanced and became predominant at low temperature. Here, in order to further understand the magnetic ground state and magnetic QCP in YFe$_2$Ge$_2$, we conduct an $^{89}$Y NMR study on the single crystals, which are from the same sample batch for the recent INS experiment\,\cite{neutron}. An incoherent-to-coherent crossover due to Hund's coupling induced electronic correlation is unambiguously revealed, which has already been observed in KFe$_2$As$_2$\,\cite{Hardy,ypwu,zhao}. Interestingly, our results also indicate that the low-temperature enhancement of A-type spin fluctuations observed by the INS experiment is tightly bounded to the low-temperature coherent state. Therefore, we conclude that, below the crossover temperature, the YFe$_2$Ge$_2$ system is approaching an itinerant magnetic QCP. Our results shed new light on understanding the correlation between AFM and FM spin fluctuations in FeSCs.

\begin{figure}[!t]
	\centering
	\includegraphics[width=0.9\columnwidth,angle=0,clip=true]{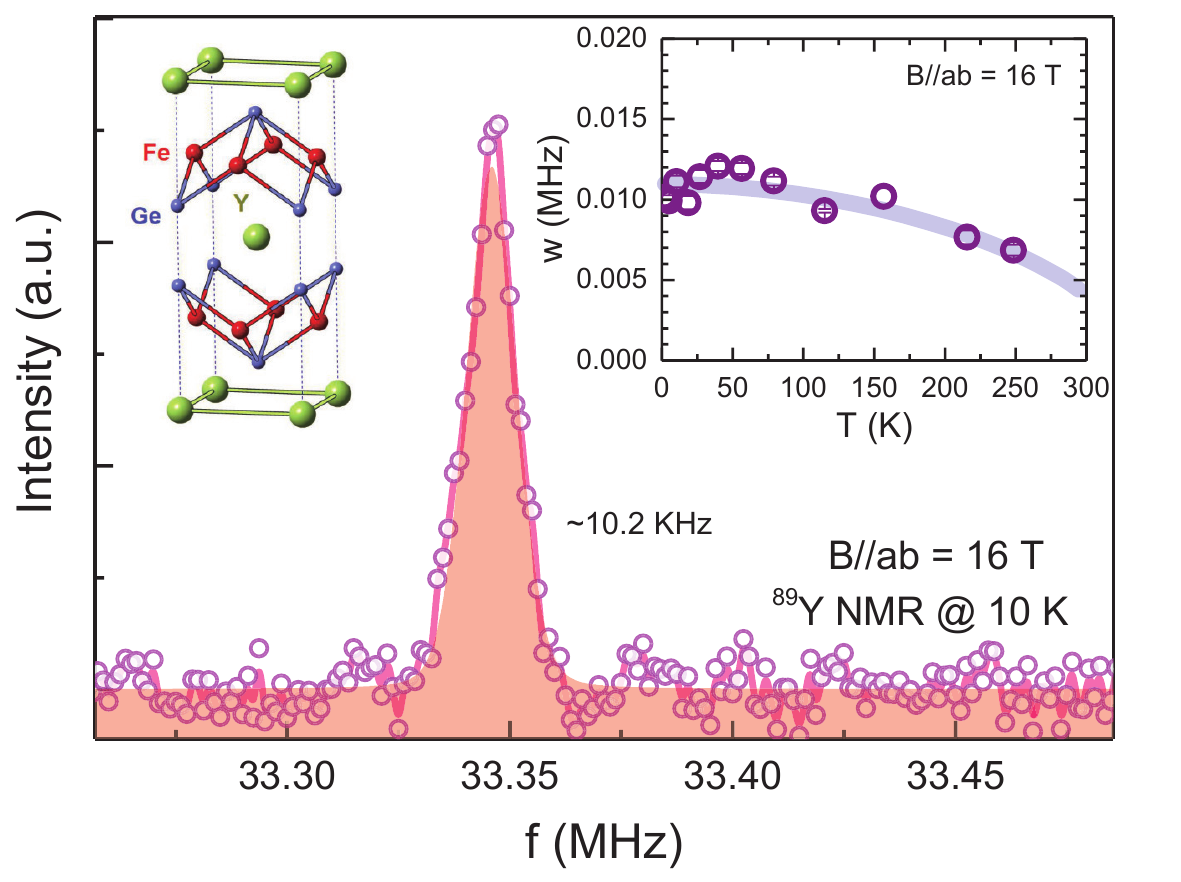}
	\caption{(Color online) The full spectrum of $^{89}$Y at 3.2 K with external field parallel to the $ab$ plane. The crystal structure of YFe$_2$Ge$_2$ as shown in the inset. Inset: The temperature-dependent linewidth of the $^{89}$Y NMR spectrum with the external field along the $ab$ plane. }
	\label{fig1}
\end{figure}
\section{II. METHOD}
High-quality YFe$_2$Ge$_2$ single crystals were synthesized by the tin-flux method\,\cite{neutron}. The present NMR measurement on $^{89}$Y nuclei is conducted from 2 to 300 K. The external magnetic field of 16 T is applied parallel to either the $c$ axis or the $ab$ plane. As shown in Fig. 1, the linewidth shows a weak temperature dependence and is $\sim10\,\mathrm{KHz}$ at low temperature. Compared with previous NMR results on polycrystalline samples\,\cite{YFe2Ge2_NMR}, this narrow linewidth indicates that the single crystal used in the present NMR study is of very high quality.

\begin{figure}[!t]
	\centering
	\includegraphics[width=0.9\columnwidth,angle=0,clip=true]{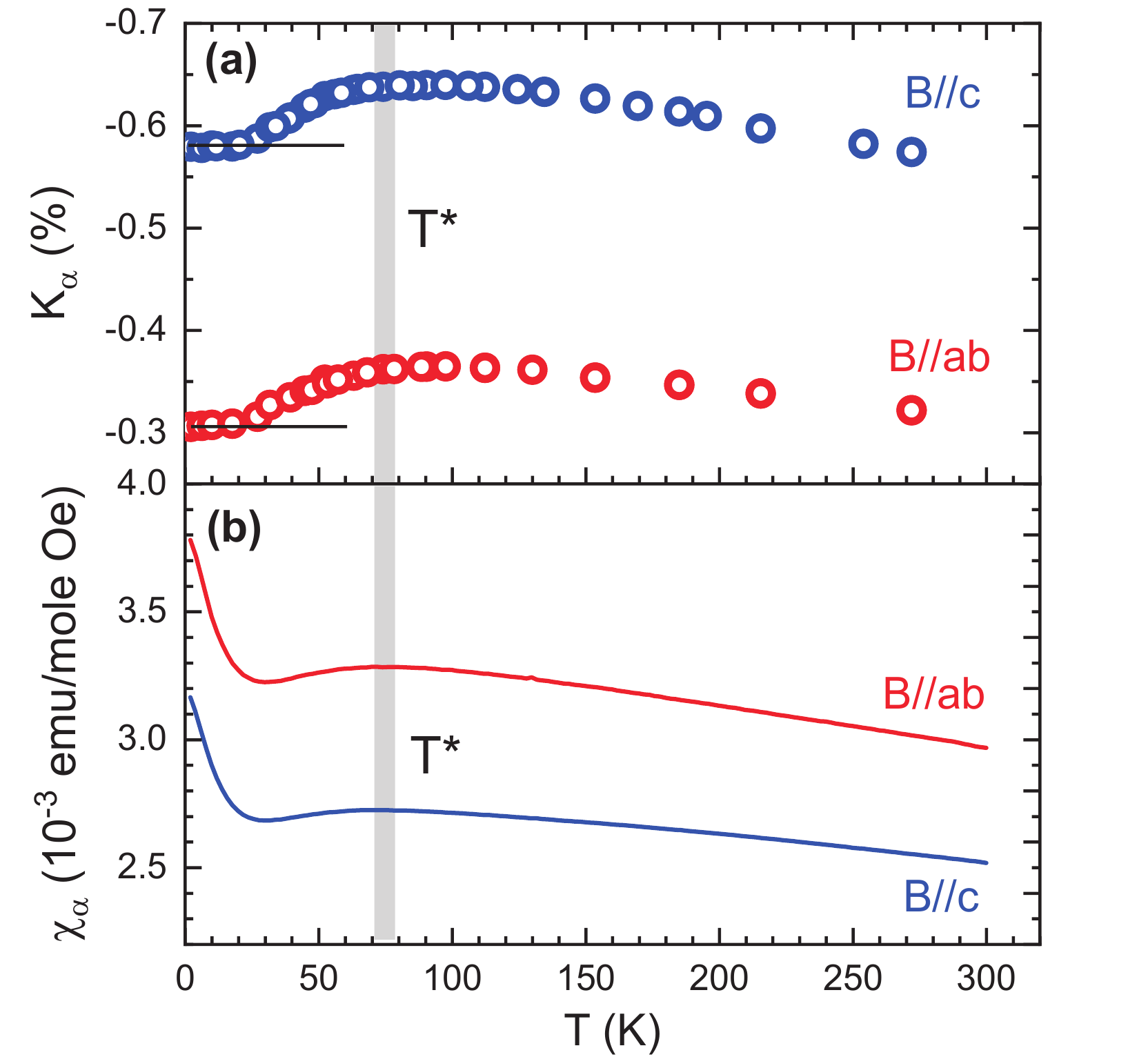}
	\caption{
		(Color online) (a) Temperature-dependent Knight shift of $^{89}$Y with the field parallel to the $ab$ plane and $c$, receptively. The electronic crossover temperature is indicated by the gray bold line with $T^*\sim75\pm15\,\mathrm{K}$. (b) Bulk magnetic susceptibility ($\chi$) of YFe$_2$Ge$_2$ versus temperature ($T$) with an external field of 5 T along the $ab$ plane and $c$ axis, respectively.
			     }
	\label{fig2}
\end{figure}
\section{III. RESULTS}
First, clear experimental evidence for electronic crossover behavior is observed in YFe$_2$Ge$_2$ by both bulk magnetic susceptibility and Knight shift measurement. In general, the temperature-dependent Knight shift can be expressed as $K(T)=K_{orb}+A_{hf}\chi_{bulk}(T)$, where $K_{orb}$ is a $T$-independent orbital shift, $A_{hf}$ is the hyperfine coupling tensor between nuclear spins and electron spins, and $\chi_{bulk}$ is the uniform spin susceptibility. When there is only one spin degree of freedom, the Knight shift $K(T)$ can be scaled with the bulk susceptibility $\chi_{bulk}(T)$ and both of them show a similar temperature dependence. As shown in Fig. 2(a), the Knight shift of $^{89}$Y exhibits a similar electronic crossover behavior as that in \emph{A}Fe$_2$As$_2$ (\emph{A}=K, Rb, and Cs) family\,\cite{ypwu}. Above $T^*\sim75\pm15\,\mathrm{K}$, the Knight shift $K(T)$ increases with decreasing temperature, which is consistent with the bulk magnetic susceptibility $\chi_{bulk}$ in Fig. 2(b). Such temperature-dependent behavior in both $K(T)$ and $\chi_{bulk}(T)$ served as evidence of local moments\,\cite{Hardy,ypwu}. Below $T^*$, the Knight shift gradually decreases with further lowering temperature and then becomes saturated below 16 K. Except for a Curie-tail behavior at low temperature, the bulk magnetic susceptibility is quite consistent with the Knight shift below $T^*$, supporting such electronic crossover behavior. The Curie-tail behavior in the bulk magnetic susceptibility is usually ascribed to the impurity effect, which would only affect the NMR linewidth but not for the Knight shift. Therefore, the nearly $T$-independent $K(T)$ at low temperature is related to an intrinsic uniform spin susceptibility, which suggests a coherent state with a Pauli-like paramagnetism. A similar coherent state is also observed in \emph{A}Fe$_2$As$_2$ (\emph{A}=K, Rb, and Cs)\,\cite{ypwu,zhao,Hardy}.

In order to further understand the nature of such electronic crossover, we carefully check the quantitative relation between the Knight shift and the bulk magnetic susceptibility in YFe$_2$Ge$_2$. As shown in Fig. 3, the $K$-$\chi_{bulk}$ plot exhibits a clear deviation from the high-temperature linear behavior at $T^*$. Such nonlinear behavior is usually called a Knight shift anomaly, which is due to the existence of multiple spin degrees of freedom\,\cite{curro}. Considering multiple spin degrees of freedom, the total Knight shift can be rewritten as $K(T)=K_0+A_1\chi_1(T)+A_2\chi_2(T)+\cdots$, while the total spin susceptibility is expressed as $\chi(T)=\chi_1(T)+\chi_2(T)+\cdots$. If $A_1=A_2=\cdots$, then $K(T)$ can be still scaled with $\chi(T)$. If $A_1\neq A_2\neq\cdots$ and each spin susceptibility component $\chi_i(T)\,(i=1,2,\ldots)$ also has different temperature dependence, then $K(T)$ will no longer be scaled with $\chi(T)$. This is called the Knight shift anomaly. Therefore, the emergent Knight shift anomaly below $T^*$ indicates that multiple spin degrees of freedom are involved in the electronic crossover of YFe$_2$Ge$_2$. In order to exclude the possible origin from the impurity effect, we further check the Knight shift anomaly in the $K_{ab}$-$K_c$ plot, in which a similar nonlinear behavior is also expected for the Knight shift anomaly (see the Supplemental Material for details\,\cite{Supplemental}). As shown in the inset of Fig. 3, a clear Knight shift anomaly is unambiguously confirmed around $T^*$, supporting the intrinsic nature of multiple spin degrees of freedom in YFe$_2$Ge$_2$.

\begin{figure}[!t]
	\centering
	\includegraphics[width=0.9\columnwidth,angle=0,clip=true]{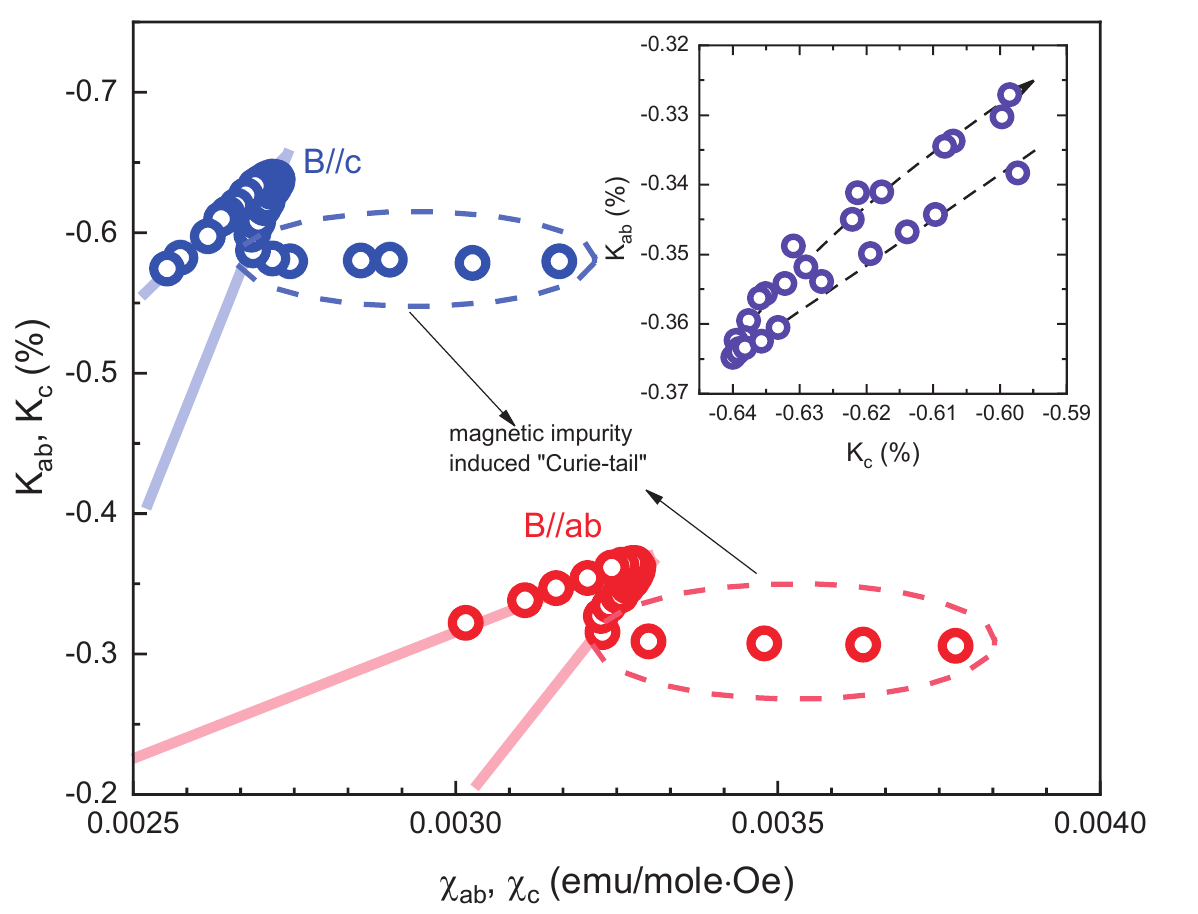}
	\caption{
				(Color online) $K$-$\chi$ plot for YFe$_2$Ge$_2$. Due to the magnetic impurities effect, the data points circled by the dotted line do not reflect the intrinsic property of YFe$_2$Ge$_2$. Inset: $K_{ab}$-$K_c$ plot for YFe$_2$Ge$_2$. The arrow direction is from high temperature to low temperature. The inflection point is around $T^*$.
	}
	\label{fig3}
\end{figure}

The similar electronic crossover and Knight shift anomaly have already been observed in \emph{A}Fe$_2$As$_2$ (\emph{A}=K, Rb, and Cs) family\,\cite{Hardy, ypwu}, which are ascribed to an incoherent-to-coherent crossover due to the Hund's coupling\,\cite{Hardy}. In this picture, the Hund's coupling induced orbital-selective electronic correlation can naturally explain the multiple spin degrees of freedom suggested by the Knight shift anomaly\,\cite{ypwu, zhao}. The great similarity of the electronic crossover behavior between YFe$_2$Ge$_2$ and \emph{A}Fe$_2$As$_2$ (\emph{A}=K, Rb, and Cs) strongly suggests that a similar physical scenario is also suitable for YFe$_2$Ge$_2$. This also qualifies the YFe$_2$Ge$_2$ as a reference system to understand \emph{A}Fe$_2$As$_2$ (\emph{A}=K, Rb, and Cs) family. In addition, the similar temperature-dependent behavior of $K(T)$ was also observed in the early NMR measurement on a polycrystalline sample\,\cite{YFe2Ge2_NMR}.

\begin{figure*}[!t]
	\centering
	\includegraphics[width=0.9\textwidth,angle=0,clip=true]{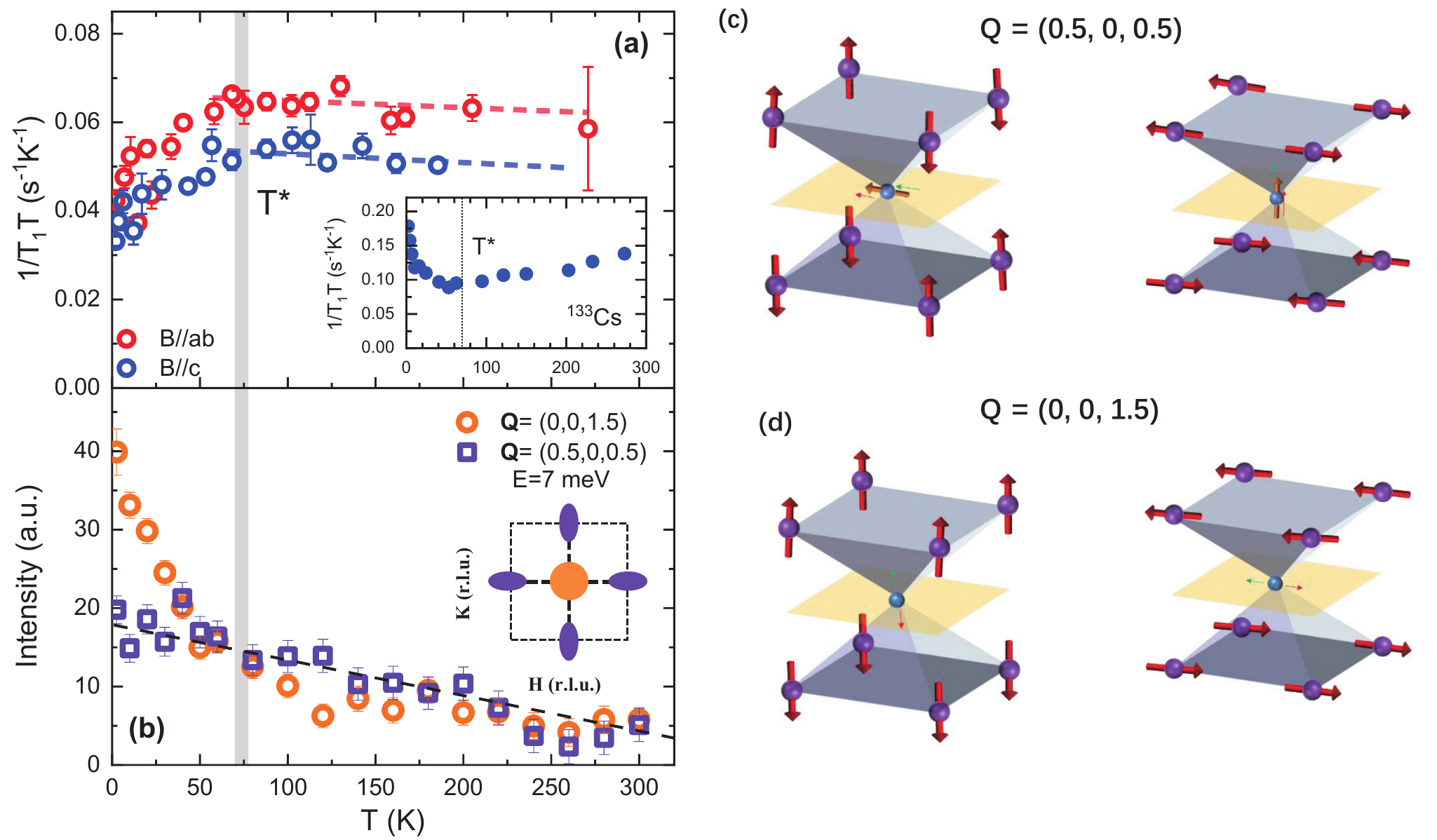}
	\caption{
	    (Color online) (a) $1/T_1T$ versus temperature with an external field of 16 T parallel to the $ab$ plane and $c$ axis, respectively. The inset shows the $1/T_1T$ of $^{133}$Cs in CsFe$_2$As$_2$.
		(b) Temperature dependence of the A-type spin fluctuations with $\chi^{''}(E=7\,\mathrm{emV},\mathbf{Q}=(0,0,1.5))$ and stripe-type spin fluctuations with $\chi^{''}(E=7\,\mathrm{emV},\mathbf{Q}=(0.5,0,0.5))$. The data are taken from the previous INS measurement\,\cite{neutron}.
		(c), (d) Schematic illustration of the transferred hyperfine fields at $^{89}$Y sites. The sources of the hyperfine field comes from stripe-type spin fluctuations (0.5,0,0.5) as shown in (c). The sources of the hyperfine field comes from A-type spin fluctuations (0,0,1.5) as shown in (d).
		Violet spheres represent the Fe atoms and small blue spheres represent $^{89}$Y nuclei. The red bold arrows represent the magnetic moment on Fe sites.
	    The green and red thin arrows represent the hyperfine fields from upper and lower Fe-Ge planes respectively.
	   	}
	\label{fig4}
\end{figure*}

On the other hand, previous studies indicated that YFe$_2$Ge$_2$ is close to a magnetic QCP with a predominate in-plane ferromagnetic correlation\,\cite{Singh}. In order to further study the critical spin fluctuations due to magnetic QCP in YFe$_2$Ge$_2$, we measured the temperature-dependent spin-lattice relaxation rate $1/T_1$ of $^{89}$Y nuclei. In general, the $1/T_1$ can be expressed in terms of the imaginary part of the dynamic spin susceptibility, $\mathrm{Im}[\chi(\omega_N,\mathbf{q})]$, as
\begin{equation}
	\frac{1}{T_1}=\lim_{\omega_N\to  0}\frac{\gamma^2_N}{2N}k_BT\sum_{\alpha,\mathbf{q}}F_\alpha(\mathbf{q})\frac{\mathrm{Im}[\chi^{\alpha\alpha}(\omega_N,\mathbf{q})]}{\hbar\omega_N},
\end{equation}
where the sum is over the wave vector $\mathbf{q}$ within the first Brillouin zone. $\mathrm{Im}[\chi^{\alpha\alpha}(\omega_N,\mathbf{q})]$ is the imaginary part of the dynamic spin susceptibility of electrons at the wave vector $\mathbf{q}$ and with the Larmor frequency $\omega_N$. $F_\alpha(\mathbf{q})$ is the $\mathbf{q}$-dependent form factor, which is a function of the hyperfine coupling tensor $\mathbf{A(q)}$\,\cite{angle nmr,moriya NMR}. As shown in Fig. 4(a), the temperature-dependent $1/T_1T$ slightly increases with decreasing temperature above $T^*$ and then shows a clear decrease below $T^*$. The whole temperature dependence of $1/T_1T$ is quite consistent with the temperature dependence of the Knight shift. To simplify the discussion, the spin dynamic susceptibility can be understood as $\chi^{\alpha\alpha}(\omega,\mathbf{q})=\chi^{\alpha\alpha}_{FL}(\omega,\mathbf{q})+\chi^{\alpha\alpha}_{AF}(\omega,\mathbf{q})$, where $\chi_{FL}$ stands for a weakly $\mathbf{q}$-dependent contribution as conventional Fermi liquid and $\chi_{AF}$ stands for a strongly $\mathbf{q}$-dependent contribution from the critical spin fluctuations at some $\mathbf{q}$-vector\,\cite{MMP, Monien}. When only considering the Fermi-liquid-like contribution, the $1/T_1T$ would roughly follow a similar temperature-dependent behavior as the Knight shift due to the well-known Korringa relation\,\cite{MMP}. However, as the critical spin fluctuations at a certain $\mathbf{q}$-vector come in, the $1/T_1T$ would be enhanced and break the Korringa relation. When the contribution from the critical spin fluctuations dominates, the temperature-dependent behavior of the $1/T_1T$ would be different from the Knight shift. Therefore, our above results on $1/T_1T$ suggests the absence of a contribution from critical spin fluctuations. This seems to be inconsistent with the proposed magnetic QCP scenario\,\cite{Singh}. A possible explanation for this discrepancy is to consider the filtering effect of the form factor $\mathbf{F(q)}=0$. Through a careful analysis of the form factor at $^{89}$Y sites\,\cite{Supplemental}, we found that the hyperfine field due to the A-type spin fluctuations with $\mathbf{q}=(0,0,1.5)$, which are proved as the predominant spin fluctuations by a recent INS experiment\,\cite{neutron}, is completely canceled with $\mathbf{F}(\mathbf{q})$ = 0 as shown in Fig. 4(d). Therefore, the absence of contribution from A-type spin fluctuations in $1/T_1T$ can be ascribed to such a filtering effect. A similar filtering effect of spin fluctuations has also been observed in cuprates, such as $^{89}$Y and $^{17}$O NMR in YBCO\,\cite{Alloul, YBCO_filter}. By further comparing to the INS results (as shown in Fig. 4), we found that the remarkable enhancement of A-type spin fluctuations perfectly coincides with the reduction of $1/T_1T$ below $T^*$. It means that the electronic crossover around $T^*$ drives the system approaching a magnetic QCP with a predominant A-type spin fluctuation. In addition, as suggested by previous INS experiments, besides the predominant A-type spin fluctuations, there is also a minor stripe-type spin fluctuation with $\mathbf{q}=(0.5,0,0.5)$ in YFe$_2$Ge$_2$. As shown in Fig. 4(c), there is no filtering effect on the stripe-type spin fluctuations. So the minor stripe-type spin fluctuations should contribute to $1/T_1T$. By analyzing the anisotropy of $1/T_1T$, we have successfully identified the expected stripe-type spin fluctuations (see the details in the Supplemental  Material\,\cite{Supplemental}).

\section{IV. DISCUSSION}
Next, we would like to compare the temperature dependence of $1/T_1T$ between YFe$_2$Ge$_2$ and CsFe$_2$As$_2$. The previous studies indicated that the \emph{A}Fe$_2$As$_2$ (\emph{A} = K, Rb, and Cs) family also approaches to a magnetic QCP\,\cite{CsFe2As2_London_penetration_QCP,CsFe2As2_strain_QCP}. Previous INS experiments on KFe$_2$As$_2$ found that the predominant AFM spin fluctuations in this family are located at $\mathbf{q}=[\pi(1\pm\delta),0]$ with $\delta=0.16$\,\cite{neutron_KFe2As2}, which will not suffer the filtering effect at the interlayer Cs sites. As shown in the inset of Fig. 4(a), the temperature-dependent $1/T_1T$ at $^{133}$Cs sites shows that a remarkable enhancement of $1/T_1T$ emerges just below the incoherent-to-coherent temperature with $T^*\sim75\mathrm{K}$\,\cite{ypwu, zhao}. This result indicates that the remarkable enhancement of spin fluctuations in CsFe$_2$As$_2$ is also driven by the electronic crossover around $T^*$ as that in YFe$_2$Ge$_2$. Both of these facts indicate that the enhanced spin fluctuations below $T^*$ are actually related to an emergent coherent state. In this sense, the magnetic QCP in these systems should exhibit an itinerant nature. The previous DFT calculations have successfully predicted the critical spin fluctuations in both YFe$_2$Ge$_2$ and the \emph{A}Fe$_2$As$_2$ family from an itinerant picture\,\cite{Singh, neutron_KFe2As2, interlayer cation}. This is also consistent with our present conclusions. Considering the Hund's coupling induced electronic correlation in these systems, the itinerant picture is not necessary to be correct. A local spins model has also been proposed for understanding the magnetic QCP in \emph{A}Fe$_2$As$_2$ (\emph{A} = K, Rb, and Cs)\,\cite{CsFe2As2_strain_QCP}. So why does the itinerant picture work so well in these systems? The key point is the Hund's coupling induced incoherent-to-coherent crossover, which has a very similar role as the Kondo crossover in heavy fermion systems\,\cite{Yifeng_Yang}. In heavy fermion systems, the nature of magnetic QCP (local or itinerant) also strongly depends on the Kondo crossover\,\cite{Gegenwart}. When the magnetic QCP is located inside the Kondo crossover, it is always itinerant in nature, the same as YFe$_2$Ge$_2$ and \emph{A}Fe$_2$As$_2$ (\emph{A} = K, Rb, and Cs). In addition, a similar correlation between FM spin fluctuations and electronic crossover was also observed in Sr$_2$RuO$_4$, in which the entire electronic system also develops into a coherent state accompanied by the growth of low-energy FM spin fluctuations in the RuO$_2$ plane\,\cite{Sr2RuO4_Imai}. The Hund's coupling induced orbital-selective electronic correlation also plays a key role in this case, suggesting a universal picture among all these materials\,\cite{Sr2RuO4_orbital_theory}.

On the other hand, after the confirmation of the A-type spin fluctuations with in-plane FM correlation in YFe$_2$Ge$_2$, a natural question is how to understand the interplay between the in-plane FM spin fluctuations and superconductivity in YFe$_2$Ge$_2$. The previous angle-resolved photoemission spectroscopy result suggests that the electron-phonon coupling should be taken into account for the pairing mechanism in YFe$_2$Ge$_2$\,\cite{ARPES}. If the superconductivity is really induced by electron-phonon interaction, then it will be strongly suppressed by the low-temperature predominant FM spin fluctuations in the frame of conventional theory\,\cite{FM_SC,FM_SC_a,FM_SC_b}, which might be inconsistent with the low $T_c$ in this system. An alternate scenario to the electron-phonon picture is spin-fluctuation-mediated superconducting pairing. In general, the FM spin fluctuations favor the spin-triplet pairing and are incompatible with the spin-singlet pairing, while the AFM spin fluctuations behave in an opposite manner. In this case, the coexistence of AFM and FM spin fluctuations may also lead to a low $T_c$. In addition, the predominant FM spin fluctuations below $T^*$ strongly suggest the pairing mechanism in YFe$_2$Ge$_2$ might favor a spin-triplet pairing, which is consistent with previous electronic structure calculations\,\cite{Singh}. This still needs more experiments to confirm, such as a Knight shift measurement below $T_c$. Considering the similar Fermi surface geometry between YFe$_2$Ge$_2$ and the CTP of \emph{A}Fe$_2$As$_2$ (\emph{A} = K, Rb, and Cs), the enhanced FM spin fluctuations may also exist in the CTP of \emph{A}Fe$_2$As$_2$. If this is true, then the nonmonotonic behavior of $T_c$ in \emph{A}Fe$_2$As$_2$ under pressure can be related to the competition between AFM and FM spin fluctuations\,\cite{KFe2As2_pressure1,KFe2As2_pressure2,KFe2As2_pressure3}. A possible spin-triplet pairing is also expected in the CTP of \emph{A}Fe$_2$As$_2$. In conclusion, the present work indicates that YFe$_2$Ge$_2$ provides a good platform to study the relation between spin fluctuations and superconducting pairing in FeSCs. Moreover, a potential spin-triplet superconductivity may exist in both YFe$_2$Ge$_2$ and \emph{A}Fe$_2$As$_2$ under high pressure.

\vbox{}

\begin{acknowledgments}
\section{ACKNOWLEDGMENTS}
	This work is supported by the Science Challenge Project of China (Grant No. TZ2016004), the National Key R$\&$D Program of the MOST of China (Grants No. 2016YFA0300201 and No. 2017YFA0303000), the "Strategic Priority Research Program (B)" of the Chinese Academy of Sciences (Grant No. XDB25010100), the National Natural Science Foundation of China (Grants No. 11888101 and No. 11522434) and the Anhui Initiative in Quantum Information Technologies (Grant No. AHY160000).
\end{acknowledgments}

\end{document}